# Small Private Online Judge: A New Tool for Empirical Education Research

ZHU Yun-chi
School of Biological Sciences & Medical Engineering
Southeast University
Nanjing, China
213170096@seu.edu.cn

TONG Cheng-da
School of Biological Sciences & Medical Engineering
Southeast University
Nanjing, China
213170761@seu.edu.cn

ZHAO Zuo-han
School of Biological Sciences & Medical Engineering
Southeast University
Nanjing, China
213170440@seu.edu.cn

XIA Xiao-jun
School of Biological Sciences & Medical Engineering
Southeast University
Nanjing, China
xxj.rcls@seu.edu.cn

*Abstract*—This paper puts forward the concept of Small Private Online Judge (SPOJ). Compared with Massive Open Online Judge (MOOJ), SPOJ has advantages in structured data acquisition of students' virtual behavior for its specific function and tight coupling with the classroom. SPOJ-based empirical education research can be conducted within "Acquisition-Analysis-Application" (3A) Framework. The case study of a SPOJ program clarifies the standard pattern of SPOJ-based 3A research and highlights the emergence of education-intelligence concept. The challenges of SPOJ-based empirical education research and implications of SPOJ are also discussed.

## I. BACKGROUND

Online Judge (OJ) is a tool of task arrangement and learning-status-checking effective in today's programming competitions and courses. Through black-box testing on the cloud, it can make immediate judgments about the codes submitted by users and give corresponding scores and rankings, hence it has been applied in programming education and algorithm competitions for a long time [1].

OJ is not only a high-performance evaluation service platform, but also a valuable educational data acquisition tool [2]. The development of Internet technology has led to the rise of online learning and relevant educational data mining. Compared with those in traditional classrooms, students' various online learning behaviors are virtualized and structured, consequently easier to be acquired in a high-throughput way. Under such background, we propose the role of small private online judge (SPOJ) in empirical education research related to programming courses.

## II. BASIC CONCEPTS OF SPOJ

### A. Classification: MOOJ and SPOJ

Similar to Online Course, OJ can be classified into massive open online judge (MOOJ) and small private online judge (SPOJ) according to their different attributes (Table I).

TABLE I. TABLE TYPE STYLES

|  | **MOOJ** | **SPOJ** |
| --- | --- | --- |
| Accessibility | Public | Private |
| Administrator | University / Business | Teacher |
| Orientation | Algorithm competition | Programming course |
| Knowledge coverage | Board | Specific |
| Target user | Self-learner | Student |
| User & problem scale | Large & Growing | Small & Stable |
| Scoring rule [3] | ACM/OI | ACM |

Just like SPOC derives from the booming MOOC, SPOJ emerges with the development and improvement of MOOJ. The launch of the early competition-oriented MOOJs such as UVa, POJ and Codeforces laid the foundation for the development framework and service pattern of OJ system. Jointed by the growing application of recent high-quality open-source projects, such as HUSTOJ and QDUOJ, the present technology allows SPOJs to be quickly deployed and applicated to programming teaching.

Different from the MOOJs listed above, SPOJs are "Small" and "Private", for they are course-oriented OJs built by teachers aiming to provide more efficient programming task assessment for their specific classes. SPOJs are usually deployed in the campus network which cannot be accessed publicly, thus they exist in a numerous-unknown status.

## B. The data-acquisition of SPOJ

The database structure of most OJs includes the following four main parts (Figure 1):

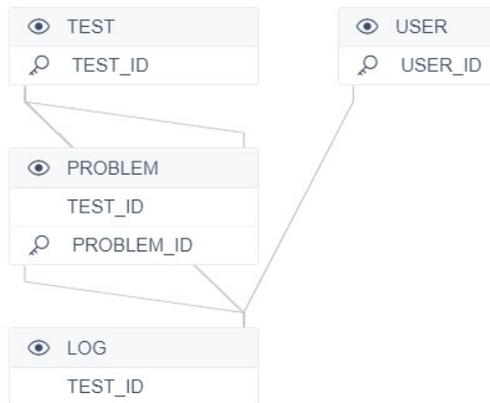

Figure 1.  OJ Data Association View.

i) LOG: Log data is the most important data in OJ's database. It reflects all the behavior of a user at a certain time on OJ, and is stored in a table named "Solution " or "Status". The basic structure of log data is illustrated in Table II:

TABLE II.  BASIC STRUCTURE OF LOG TABLE

| Name | Description | Sample |
| --- | --- | --- |
| ID | Auto increment | 1 |
| ProblemID |  | 1 |
| TestID | \ | 1 |
| UserID |  | D-4282 |
| InTime | User's submission time, usually accurate to the second | 2020-02-03 09:00:00 |
| Language | \ | C++ |
| Result | Based on the scoring rules | True |
| Time |  | 10ms |
| Memory | Details | 1M |
| CodeLength |  | 428 |
| Code |  | #include<iostream>··· |
| IP | User's IP when submitting | 47.98.171.106 |
| … | … |  |

Compared with other online systems, OJ is more function-specific in that users only submit code and view results there while specific programming operations are done locally. Although unable to capture the detailed programming behavior, OJ's log runs as an efficient programming snapshot system recording a user's continuous submission towards a problem until it is passed or abandoned. In addition, it provides teachers access to students' usage habits from the log, such as their usage time, places of study*, etc.

ii) PROBLEM: The problem data mainly includes ID, title, description, sample input, sample output, etc. Each problem requires at least one set of test data which is not usually written to the database yet.

The quantity of SPOJ's problems, compared with that of MOOJ, is much smaller and relevantly stable, and tends to be adjusted more frequently due to teaching reform.

iii) TEST: Test refers to the set of time-limited problems released by teacher on SPOJs in the form of exams or assignments for students. The test data includes ID, start and end time, etc.

Different from the public problem list of MOOJ, problems on SPOJ are usually posted in tests. To some extents, problems released from teachers are SPOJ's input signals while logs generated by students' virtual behavior are its output signals, so teachers tend to utilize tests to keep all signals in order and bounded.

iv) USER: The user data usually includes only the most basic information such as ID and user name when students' detailed information has been registered on the campus authentication platform.

While sharing similar database structure and task scheduling mechanism with MOOJ, SPOJ has many advantages over it in data acquisition. Firstly, SPOJ's deployer and administrator have a closer relationship (often even the same person), thus richer types of data can be exported directly from the database instead of using web crawlers or API. Secondly, SPOJ is closely integrated with classroom-teaching where teachers' assignment release and student task completion are periodic, contributing to more regular data generation. Finally, the scale of SPOJ users is small and stable, and user behavior is much easier to control, which improves the security of SPOJ and reduces the difficulty of data cleaning.

SPOJ researchers concentrate on log data rather than problem data. For one thing, due to users' elementary programming skill and syllabus regulations for limited sphere of knowledge points, the degree of difficulty and discrimination of course-oriented OJ problems is lower and more uniformly-distributed than competition-oriented problems. Therefore, there is less need for SPOJ researchers to focus on problem data to build applications such as learning topic and difficulty level recommendation system [4]. For another, SPOJ researchers usually have a closer relationship with its administrator as well as users, so they may pay more attention to students' virtual behavior pattern to be mined from log data.

## C. 3A Framework Based on SPOJ

SPOJ-based empirical education research fits the research framework of "Acquisition-Analysis-Application" (3A) [5] illustrated in Figure 2. SPOJ is the infrastructure supporting these researches where the raw data is generated and stored. Researchers can acquire smaller-scale structured data directly from the database and use relatively few computing resources for analysis.

Different from those under traditional "hypothesis-testing" research framework [6], empirical education researches under 3A Framework aim to mine patterns from the unknown so as to develop applications to solve practical problems. Within 3A Framework, the research can lead to both pedagogical and technological improvement. The pedagogical application may include student virtual behavior norm, early-warning model fused OJ data with exam scores, and various education-intelligence reports, while the technological application may be SPOJ's management optimization, classroom-teaching improvement, etc. A complete 3A research is equivalent to the construction of an ETL pipeline connecting SPOJ with the applications from this perspective.

---

* The distribution of campus LAN IPs is usually published on the school network center's official website. The IP recorded by OJ allows teachers to observe on-campus locations where students complete the assignments, such as a computer room, library, or dormitory.

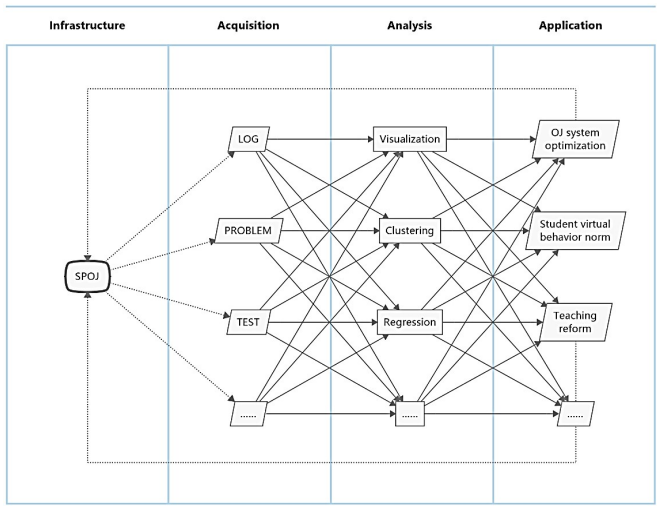

Figure 2. 3A Framework Based on SPOJ.

## III. CASE STUDY

### A. Introduction

In 2017, the course leader deployed HUSTOJ on Alibaba Cloud (http://oj.bmeonline.cn) for release and inspection of programming assignments. Every assignment was released in the form of a test of 4 to 8 problems with a deadline of 1-2 weeks. User registration was done by the students themselves, and they were allowed to complete the tests off-class or during on-machine classes scheduled by the school. 76 problems were released on the OJ via 11 tests during 3 months, and 10,491 logs were generated from 96 students [7].

In 2018, the teacher still allowed students to register and complete autonomously. The 2017 tests were reused, only with the start date reset to keep pace with teaching progress. What's more, no deadline was reset. This semester, only 8493 valid logs were generated on the OJ from the 104 recruited students, a year-on-year decrease of 19% [8].

In 2019, SEU adopted large-categories enrollment and small-class teaching. The original OJ was replaced with QDUOJ (http://www.bmeonline.cn/oj) deployed in the campus LAN, and the user-scale was reduced to 26 students from the electronic-information category. OJ's registration was changed to unified certification while the supervision of on-machine classes became much stricter. This semester, 12 tests with 68 problems were posted on the OJ, generating 2609 logs [9].

### B. Acquisition and Analysis

The school's teaching researcher, a data engineer rather than a learning scientist, began exporting and analyzing data from the SPOJ's database at the end of 2018.

For dimensionality reduction, most of the irrelevant variables were removed, including the TIME($t$) and MEMORY ($m$), which have been the criteria for judging algorithm competition problems for a long time. These variables either lacked statistical significance (Table III, taking $t$ and $m$ as examples), or had nothing to do with empirical education research.

TABLE III. RESULTS OF Z-TEST

| Year | Variable | $H_0$ | $\mu_0$ | $P_{value}$ | Result |
|---|---|---|---|---|---|
| 2017 | $t$ | | 22.89 | 0.6936 | |
|  | $m$ | $\mu=\mu_0$ | 2032 | 0.9712 | Accept $H_0$ |
| 2018 | $t$ | | 22.89 | 0.6741 | |
|  | $m$ | | 2032 | 0.9692 | |

For sample size reduction, when searching for logs unrelated to students in 2018's data, a large number of subsidiary accounts (alt) were found. It pushed the researcher to spend much time combining accounts or deleting relevant records directly, causing samples lost beyond prediction. No other abnormal user behavior was found during data cleaning.

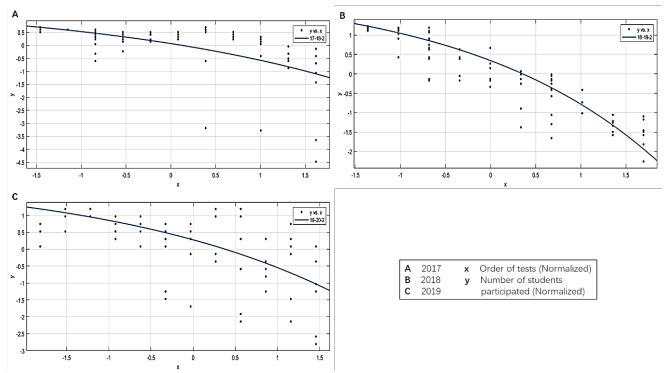

Figure 3. Participation Curve.

The 2017-2018 students' involvement with OJ problems reveals the number of participants declined exponentially over time (Figure 3 A and B). The gradually increasing difficulty of OJ problems along with classroom-teaching progress may account for the participants decrease and intra-test discrimination increase. However, the decline in 2018 appears to be too large.

The attempt to establish a correlation between students' completion of OJ problems and final course grades does not lead to positive result. The researcher defined the *AC* index according to the ACM judging rule [2], and calculated the Pearson Correlation Coefficient between each student's *AC* and their exam scores (*MSC&WSC*). All results only meet the criteria for moderate correlation (Table IV). Such moderate correlation may be caused by the difference between the capability to be examined and the capability to be trained, for students trained their practical programming skills on the OJ while the final exam focused on theoretical knowledge.

TABLE IV. CORRELATION COEFFICIENT BETWEEN *AC* AND GRADES

| Year | $\rho_{AC,MSC}$ | $\rho_{AC,WSC}$ |
| --- | --- | --- |
| 2017 | 0.7131 | 0.6074 |
| 2018 | 0.5440 | 0.7265 |
| 2019 | 0.6697 | 0.6922 |

The prominent high *AC&MSC* correlation in 2017 may be caused by a rare set of students in this school who already have a solid programming foundation before entering university. Their role as teaching assistants during on-machine classes actually multiplied the teaching resources to programming practice of 2017 course, which consequently increase the *MSC* and *AC* correlation to some extents.

Except for 2017, the correlation between *AC* and *MSC* is not even as good as *WSC*, particularly significant in 2018. The survey of the students shows majority of them have received 12 years of exam-oriented education and tend to have formed typical learning habit, and the assumption goes that the grasp of theoretical knowledge be more correlated to learning habit than practical skill. Therefore, the representation method of students' learning habits reflected in their virtual behavior on OJ is further observed.

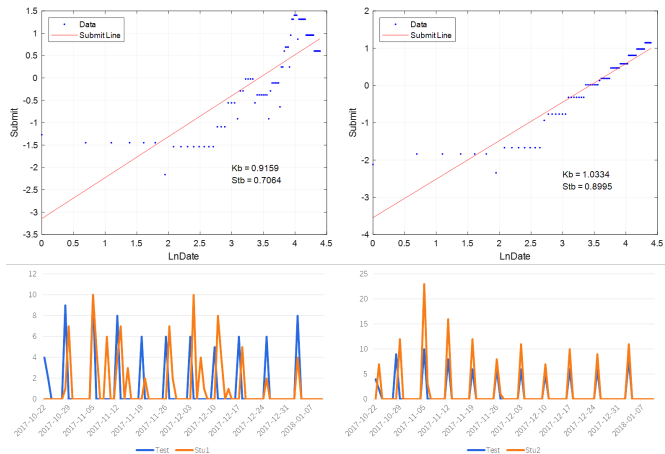

Figure 4. Submit Line Model.

Several models were established, among which *Submit Line* (Figure 4, samples from two students in 2017) [2] tends to be more interpretative. It is determined with linear regression model fitting corrected cumulative submissions against logarithmic date series of a student. This model can reflect both a student's effort in each test ($K_b$, slope of *Submit Line*) and his or her time arrangement habits for completing them ($St_b$, $R^2$ of the fit). The submit-lines in Figure 4 demonstrates that Student 2 is more devoted in OJ problems along with more timely completion than Student 1, which matches their daily submission curves below.

C. Application

Findings from above analysis are applied to two types of practice:

i) Strengthening management of OJ: Contrast between data from last two years made the course leader realized the necessity for stricter management, hence he took measures as mentioned in introduction (III.A). As a result, the downward trend of *Participation Curve* of 2019 (Figure 3 C) becomes much gentler than that of 2018, not as gentle as that of 2017 yet.

ii) Building education-intelligence (EI) system: In the analysis process, inspired by the concept of business intelligence [10], the researcher proposed "education intelligence", which means utilizing advanced data technology to build ETL pipelines within schools in a gesture to promote the 3A-cycle-flow of education data, the rationality of educational decision-making, and the improvement in education effectiveness. Guided by this concept, the researcher "employed" one of the three "teaching assistants" (III.B) to transform his analysis methods into software assisting the cleaning, visualization and analysis of OJ data. The software has been developed since the summer of 2019 and now is open-source under the MIT license [11], representing the completion of the first OJ data ETL pipeline aiding programming teaching.

D. Disccusion

The empirical education research within 3A Framework reported above has exerted a positive effect on programming teaching practice. It also revealed challenges in SPOJ-based education research:

i) Special behavior influence: SPOJ's user relationship is usually very close, for they are students learning in the same classroom and even living in the same apartment. Therefore, special behavior of individual users may have a significant impact on the user group whether positive or negative. While strong measures must be taken to eliminate the effects of negative behavior, positive behavior contributing to a virtuous circle in SPOJ user ecology is more worthy of attention, recording and researching.

ii) Black-box attribute: There is sort of "deception" in any virtual behavior. A student with a good programming foundation may not take the basic problems seriously while another student who always completes assignments in a timely and efficient manner may be a cheater actually. More data is required to accurately represent such behavior, among which source code data from SPOJ has great potential due to development of against-cheating systems. Teachers' understanding of student information and behavior-based cluster analysis also help capture such "deceptive" behavior.

iii) Internalization effect: The requirement for strict management of SPOJ makes it inaccessible to the public and restricted in user scale. This guarantees the operating efficiency and data quality of SPOJ, but is not conducive to communication between peer teachers as well as the expansion of education-intelligence. A solution for this dilemma is to establish a public database and open SPOJ data in accordance with a recognized OJ data standard. At present, Xia et al in SEU has begun to explore in this field [12].

iv) Theoretical basis: Application-oriented as it is, SPOJ-based education research still demands theoretical support from learning science, which may explain the more complex cognitive mechanisms and emotional activities behind virtual behaviors on SPOJ, making up for the "dead zone" cause by its black-box attribute from another perspective.

## IV. Implications

SPOJ is the combination of technology innovation and course improvement. It can be defined as "OJ + Classroom", while its implication is far beyond the classroom. First, the education-data-miners on SPOJ may do class-useful empirical education research under the guidance of 3A Framework, and take appropriate measures to share their findings and expand education-intelligence. Second, course teachers or leader can utilize other technologies such as SPOC and "Rain Classroom" to enrich classroom teaching while providing more data for teaching researchers. Last but not least, in nowadays information age, disciplines relevant to learning science should attach more importance to student virtual behavior on online platforms such as SPOJ, in a gesture to provide theoretical justification for the analysis and application of education data.